\documentclass[aps,prl,twocolumn,superscriptaddress,showpacs]{revtex4-1}
\usepackage{amsmath}
\usepackage[dvips,xdvi]{graphicx}
\usepackage{amssymb}
\usepackage{epsfig}

\begin{document}
\title{First-order superfluid to Mott-insulator phase transitions in spinor condensates}
\author{J. Jiang}\thanks{These authors contributed equally to this work.}
\author{L. Zhao}\thanks{These authors contributed equally to this work.}
\affiliation{Department of Physics, Oklahoma State University,
Stillwater, Oklahoma 74078, USA}
\author{S.-T. Wang}
\affiliation{Department of Physics, University of Michigan, Ann
Arbor, Michigan 48109, USA}
\author{Z. Chen}
\author{T. Tang}
\affiliation{Department of Physics, Oklahoma State University,
Stillwater, Oklahoma 74078, USA}
\author{L.-M. Duan}
\affiliation{Department of Physics, University of Michigan, Ann
Arbor, Michigan 48109, USA}
\author{Y. Liu}
\email{yingmei.liu@okstate.edu} \affiliation{Department of
Physics, Oklahoma State University, Stillwater, Oklahoma 74078,
USA}
\date{\today}

\begin{abstract}
We observe evidence of first-order superfluid to Mott-insulator
quantum phase transitions in a lattice-confined antiferromagnetic
spinor Bose-Einstein condensate. The observed signatures include hysteresis
effect and significant heatings across the phase transitions. The
nature of the phase transitions is found to strongly depend on the
ratio of the quadratic Zeeman energy to the spin-dependent
interaction. Our observations are qualitatively understood by the mean
field theory, and in addition suggest tuning the quadratic Zeeman
energy is a new approach to realize superfluid to Mott-insulator phase
transitions.
\end{abstract}

\pacs{67.85.Fg, 03.75.Kk, 03.75.Mn, 05.30.Rt}

\maketitle

A quantum phase transition from a superfluid (SF) to a Mott-insulator
(MI) was realized in a scalar Bose-Einstein condensate (BEC) trapped
by three-dimensional (3D) optical lattices around a decade
ago~\cite{Greiner02}. Marking an important milestone, this achievement
has stimulated tremendous efforts to apply highly controllable
ultracold bosonic and fermionic systems in studying condensed matter
models~\cite{StamperKurnRMP, Campbell, LatticeRMP, Ketterle2006,
Shake2007}. The SF-MI transitions have been confirmed in various
scalar BEC systems via different techniques that can efficiently
control the ratio of interatomic interactions to the mobility of
atoms~\cite{LatticeRMP,Greiner02, Chin2010, Shake2007}. One well-known
approach to simultaneously enhance interatomic interactions and
suppress atomic motion is by raising the depth of an optical
lattice~\cite{Greiner02}. Another convenient method is to manipulate
interactions with a magnetically tuned Feshbach
resonance~\cite{Chin2010}. A third technique is to control the hopping
energy of bosonic atoms by periodically shaking the
lattice~\cite{Shake2007}. Spinor BECs, on the other hand, possess an
additional spin degree of freedom, leading to a range of phenomena
absent in scalar BECs~\cite{SengstockRPP, Sengstock2011, Hexagon,
spinorLattice, firstorder, DasSarma15}. One important prediction is
the existence of the first-order SF-MI phase transitions in
lattice-trapped antiferromagnetic spinor BECs~\cite{StamperKurnRMP,
firstorder, DasSarma15, Demler2002Spinor, Krutitsky2005First,
Kimura2005Possibility,adilet}. In contrast, the phase transitions can
only be second order in scalar BECs and ferromagnetic spinor
BECs~\cite{LatticeRMP, StamperKurnRMP, Kimura2005Possibility}.

In this paper, SF-MI phase transitions are studied in sodium
antiferromagnetic spinor BECs confined by cubic optical lattices. We
observe hysteresis effect and substantial heating across the phase
transitions, which indicate the existence of meta-stable states and
associated first-order transitions. In the ground state of the spinor
BECs, the nature of the SF-MI transitions is found to be determined by
the competition between the quadratic Zeeman energy $q_{B}$ and the
spin-dependent interaction $U_{2}$. At low magnetic fields where
$U_{2}$ dominates, signatures of first-order transitions are observed.
In the opposite limit, the transitions appear to be second order and
resemble those occurring in scalar BECs. These qualitative features
are explained by our mean-field (MF) calculations. We also study the
phase transitions with an initial meta-stable state and observe
stronger heatings across all magnetic fields. Furthermore, our data
indicate that a new technique to realize SF-MI transitions is by varying $q_B$.

Similar to Refs~\cite{fisher89, spinorLattice}, we describe a
lattice-trapped $F=1$ spinor BEC with the Bose-Hubbard model in the
lowest band as follows,
\begin{align}
H =& \frac{U_{0}}{2} \sum_{i} n_{i}
(n_{i}-1)-J \! \! \sum_{\left<
i,j\right>, m_F} \! \!  b_{i,m_F}^{\dagger} b_{j,m_F}- \mu \sum_{i} n_{i} \notag \\
& +\dfrac{U_{2}}{2} \sum_{i} (\vec{S}_{i}^{2} -2 n_{i} )+
q_{B} \sum_{i,m_F} m_F^{2}n_{i,m_F}. \label{Eq:H}
\end{align}
Here $J$ is the nearest-neighbor hopping energy, $n_i = \sum_{m_F}
n_{i,m_F}$, and $n_{i,m_F}=b^\dagger_{i,m_F} b_{i,m_F}$ is the
atom number of the $m_F$ hyperfine state at site $i$. $U_0$
characterizes the spin-independent interaction, $\mu$ is the
chemical potential, and $\vec{S}_i$ is the spin operator at site
$i$~\cite{spinOperator}. $U_{2}$ is positive (negative) in $F=1$
antiferromagnetic (ferromagnetic) spinor BECs, e.g.,
$U_2\simeq0.04U_0$ in our $^{23}$Na system~\cite{Zhao2dLattice}.
By neglecting the second order term $(b_{i,m_F}^{\dagger} -\langle
b_{i,m_F}^{\dagger} \rangle ) ( b_{j,m_F} - \langle b_{j,m_F}
\rangle )$ in the hoppings and applying the decoupling MF theory,
Eq.~\eqref{Eq:H} can be reduced to a site-independent
form~\cite{Pai2008Phases, spinorLattice, wagner2012spinor},
\begin{align}
H_{\text{MF}} = &  \dfrac{U_{0}}{2} n (n-1) +
\dfrac{U_{2}}{2} (\vec{S}^{2} -2 n)+ q_{B} \sum_{m_F} m_F^{2} n_{m_F} \notag \\
& - zJ \sum_{m_F} (\phi_{m_F}^{*} b_{m_F}  + \phi_{m_F}
b_{m_F}^{\dagger} ) + zJ |\vec{\phi}|^{2}  -\mu n
\label{Eq:HamSingleSite}
\end{align}
with the vector order parameter being $\phi_{m_F} \equiv \langle
b_{m_F} \rangle $ and $z$ being the number of nearest neighbors.
With spatially uniform superfluids in equilibrium, one can assume
$\phi_{m_F}$ to be real. $\phi_{m_F}=0$ ($\neq 0$) in the MI (SF)
phase.
\begin{figure}[t]
\includegraphics[width=85mm]{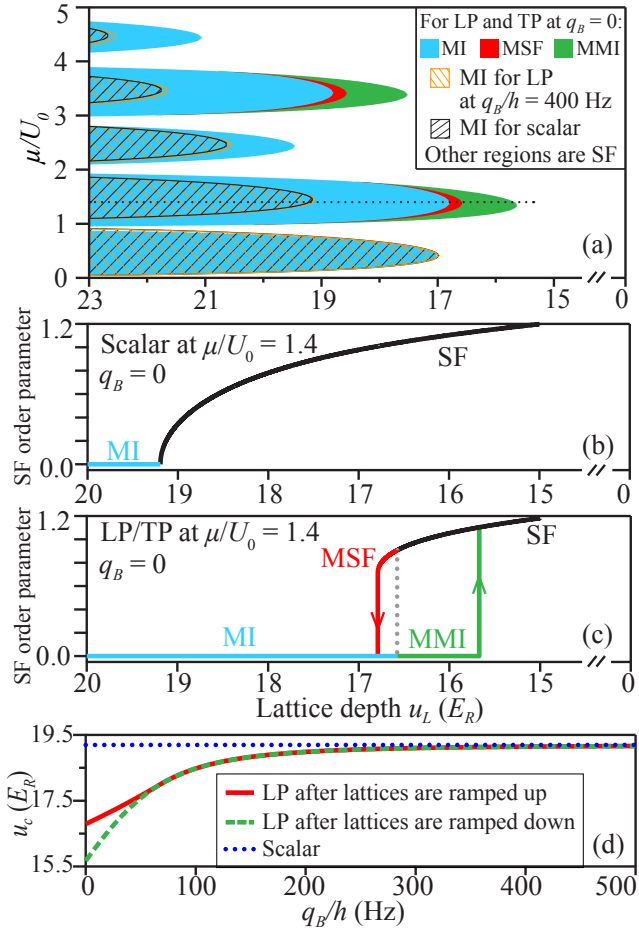}
\caption{(Color online) (a) MF phase diagrams derived from the
Bose-Hubbard model for scalar BECs~\cite{fisher89}, and the LP and TP
sodium spinor BECs in cubic lattices (see
Eq.~\eqref{Eq:HamSingleSite}). The SF order parameter versus
$u_L$ in (b) scalar and (c) LP/TP spinor BECs at $\mu/U_0=1.4$, i.e.,
along the dotted line in Panel(a). Note that SF-MI transitions are
second order in a scalar BEC, and they are first order showing
hysteresis effect in LP and TP spinor BECs at $\mu/U_0=1.4$ and
$q_B=0$. (d) Predicted SF-MI transition point $u_c$ versus $q_B$ after
cubic lattices are ramped up and down at $\mu/U_0=1.4$
(see Eq.~\eqref{Eq:HamSingleSite}).} \label{theory}
\end{figure}

An antiferromagenetic $F=1$ spinor BEC of zero magnetization forms a
polar superfluid in equilibrium with $\langle\vec S\rangle=0$~\cite{StamperKurnRMP,
JiangGS, Ho1998Spinor}. There are two types of polar superfluids: the
longitudinal polar (LP) state with $(\phi_{1}, \phi_{0}, \phi_{-1}) =
\sqrt{\rho_{s}} (0,1,0)$ and the transverse polar (TP) state with
$(\phi_{1}, \phi_{0}, \phi_{-1}) = \sqrt{\rho_{s}/2} (1,0,1)$, where
$\rho_{s}$ is the number of condensed atoms per site. At
$q_{B}=0$, TP and LP states are degenerate in energy when they have
the same $\rho_{s}$. At $q_{B}>0$, the MF ground state is always the
LP state, although a meta-stable TP phase may also
exist~\cite{StamperKurnRMP, JiangGS}. We solve
Eq.~\eqref{Eq:HamSingleSite} self-consistently by requiring
$\phi_{m_F} = \langle b_{m_F} \rangle $ in the occupancy number basis
with a maximum of 15 atoms per site. Since the observed peak occupancy
number is around six, the truncation errors are negligible.

Our MF calculations show that $q_{B}/U_{2}$ is a key factor to
understand the nature of SF-MI transitions in antiferromagenetic
spinor BECs. At low magnetic fields (where $0\leq q_{B} \lesssim
U_{2}$), $U_{2}$ penalizes high-spin configurations and enlarges
the Mott lobes for even number fillings as atoms can form spin
singlets to minimize the energy. Meta-stable Mott-insulator (MMI)
and meta-stable superfluid (MSF) phases emerge due to the spin
barrier, and lead to first-order SF-MI phase transitions (see
Figs.~\ref{theory}(a) and \ref{theory}(c))~\cite{Demler2002Spinor,
Krutitsky2005First, Kimura2005Possibility}. When 3D lattices are ramped up and down, hysteresis is expected across
the phase transitions (i.e., different transition lattice depth
$u_c$). In addition, when the system changes from a meta-stable
phase to a stable phase (e.g., from a MSF phase to a MI phase),
there will be a jump in the order parameter and the system energy, leading to unavoidable heating to the atoms.
Hence, hysteresis and substantial heating may be interpreted as
signatures of first-order transitions. As $q_B$
increases, the $m_{F}=0$ state has lower energy than other $m_{F}$
levels and $U_{2}$ becomes less relevant. When $q_B$ becomes
sufficiently larger than $U_2$ ($U_{2}/h\lesssim 80\,$Hz in this
work), the ground state phase diagram of antiferromagnetic spinor
BECs reverts back to one that is similar to the scalar
Bose-Hubbard model with only second-order SF-MI transitions (see
Figs.~\ref{theory}(a),~\ref{theory}(b) and~\ref{theory}(d)).
\begin{figure}[t]
\includegraphics[width=80mm]{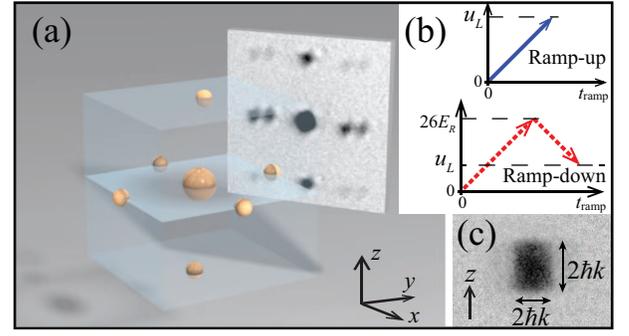}
\caption{(Color online) (a) Schematic of the reciprocal lattice
and a TOF image taken after lattices are abruptly released. This
TOF image is oriented such that its plane is orthogonal to the
imaging light. (b) Two lattice ramp sequences used in this
paper~\cite{ramp}. (c) A TOF image showing the first Brillouin zone.}
\label{schematic}
\end{figure}

Three different types of BECs (i.e., scalar BECs, LP and TP spinor
BECs) are studied in this work. A scalar BEC containing up to
$1.2\times 10^5$ sodium atoms in the $|F=1,m_F=-1 \rangle$ state is
created with an all-optical approach (see Ref.~\cite{ZhaoUwave}). A
$F=1$ spinor BEC of zero magnetization is then produced by
imposing a resonant rf-pulse to the scalar BEC at a fixed $q_B$. Since
the LP state is the MF ground state, it can be prepared by simply
holding the spinor BEC for a sufficiently long time at high magnetic
fields~\cite{JiangGS}. A different approach is required to generate
the TP state: we apply a resonant microwave pulse to transfer all
$m_F=0$ atoms in the $F=1$ spinor BEC to the $F=2$ state, and then
blast away these $F=2$ atoms with a resonant laser pulse. After
quenching $q_B$ to a desired value, we adiabatically load the BEC into
a cubic lattice by linearly raising the lattice depth $u_L$ within
time $t_{\rm ramp}$~\cite{ramp}. Lattice ramp-up and ramp-down
sequences are shown in Fig.~\ref{schematic}.
\begin{figure}[t]
\includegraphics[width=85mm]{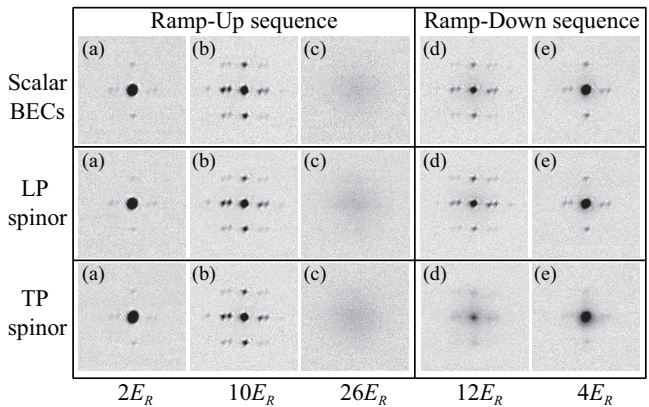}
\caption{Interference patterns observed after we abruptly release
scalar (top), LP spinor (middle), and TP spinor BECs (bottom) at
various $u_L$ and a 5.5-ms TOF at $q_B/h=360\,$Hz. Panels (a)-(c)
are taken after ramp-up sequences to a final $u_L=2, 10$, and
$26E_R$, respectively. Panels (d)-(e) are taken after ramp-down
sequences to a final $u_L$ of $12E_R$ and $4E_R$. The field of
view is $400\, \mu \text{m} \times 400 \, \mu$m.} \label{cubicTOF}
\end{figure}

\begin{figure}[t]
\includegraphics[width=85mm]{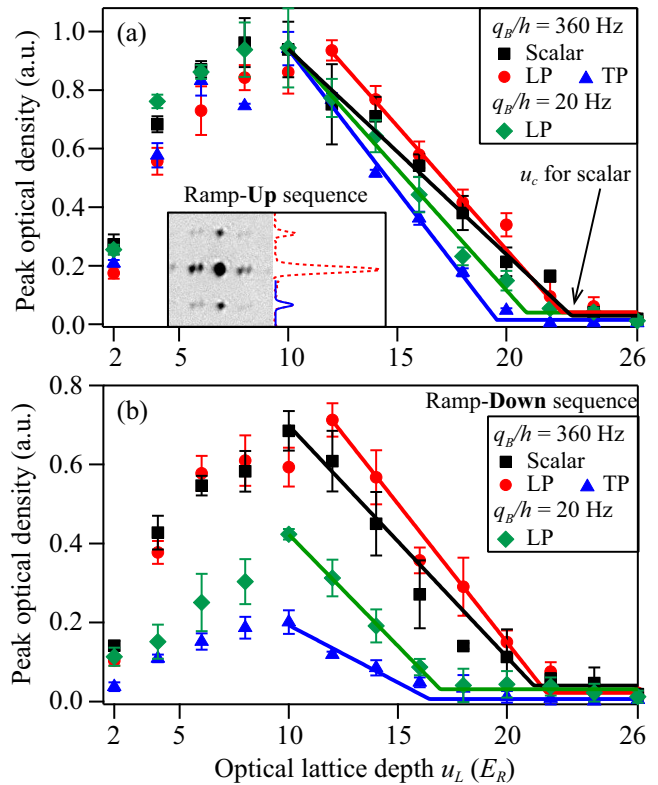}
\caption{(Color online) (a) Peak OD of interference peaks as a
function of $u_L$ after lattice ramp-up sequences. Markers are
experimental data and lines are linear fits. The critical depth $u_c$ is estimated from
the intersection of two linear fits to the data. The inset shows how we extract
the peak OD from a TOF image (left). The dotted line in the right
inset is a density profile of this TOF image through the central and
one pair of interference peaks along the vertical direction, while the
solid line is a bimodal fit to one side peak. (b) Similar to Panel(a)
except that all data are taken after lattice ramp-down sequences.}
\label{cubicDepth}
\end{figure}

\begin{figure}[t]
\includegraphics[width=85mm]{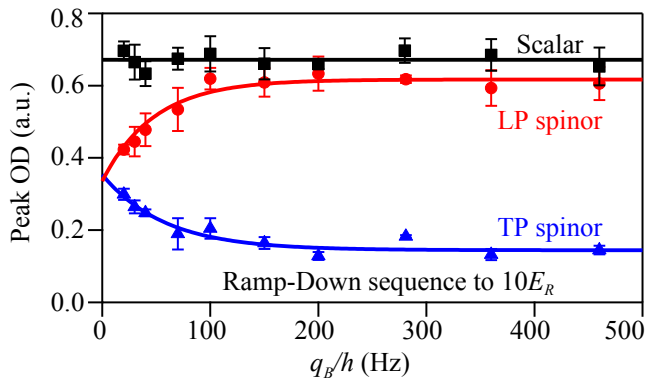}
\caption{(Color online) Peak OD of interference peaks as a
function of $q_B$ observed after lattice ramp-down sequences to
$10E_R$. Markers are experimental data. Red and blue lines
are exponential fits. The black line is a linear fit.}
\label{cubicQB}
\end{figure}

Distinct interference peaks can always be observed during
ballistic expansion, after each of the three types of BECs is
abruptly released from a shallow lattice of $u_L\leq10E_R$. Here
$E_R=h^2 k_L^2/(8 \pi^2 M)$ is the recoil energy, $M$ and $h$ are
respectively the atomic mass and the Planck constant, and $k_L$ is
the lattice wave-vector (see Ref.~\cite{KD}). As shown in the time
of flight (TOF) images in Fig.~\ref{schematic}(a) and
Fig.~\ref{cubicTOF}, the six first-order diffracted peaks are
symmetrically set apart from the central peak by a distance
corresponding to a momentum of $2\hbar k_L$ along three orthogonal
axes. These interference peaks may be considered as an indicator
for coherence associated with the SF phase in the system. In fact, a
larger visibility of interference patterns, a narrower width of
the central peak, and a higher optical density (OD) of
interference peaks have all been used as trustworthy evidence for
improved phase coherence in atomic systems~\cite{Greiner02,
LatticeRMP, Ketterle2006, Ho2007}. As an example, the LP spinor
BECs studied in Fig.~\ref{cubicTOF} demonstrate long-range phase
coherence at $u_L=10 E_R$ with a coherence length of around nine
lattice sites, which is estimated from the ratio of the central
peak width to $4\hbar k_L$~\cite{LatticeRMP, Ketterle2006}.

Figure~\ref{cubicTOF} displays TOF images at five representative
$u_L$, showing the loss and revival of the interference contrast
in both scalar and spinor BECs as cubic lattices are ramped up and
down. A quantitative analysis of these TOF images is presented in
Fig.~\ref{cubicDepth}, demonstrating that the interference peaks
(i.e., coherence associated with the SF phase) change in a reversible manner
with $u_L$. First, the interference patterns become more visible
as the lattice is made deeper, and reach their peak OD around
$10E_R$. This may be due to lattice-enhanced density
modulation~\cite{qd, LatticeRMP, Ketterle2006}. Second, when $u_L$
is further increased and exceeds a critical value $u_c$, the
interference peaks steadily smear out to a single broad peak
indicating atoms completely lose phase coherence. We read off the
value of $u_c$ in Fig.~\ref{cubicDepth} from the intersection of
two linear fits applied to the data of a given BEC. The loss of
coherence can be accounted for by many mechanisms, such as
heating, inelastic collisions, or entering into a MI state. To
confirm the system has undergone a SF-MI transition, we monitor
the lattice ramp-down sequence, because one characteristic of a MI
state has proven to be a loss of phase coherence in deep lattices
and a subsequent rapid revival of coherence as $u_L$ is
reduced~\cite{Greiner02, LatticeRMP, Ketterle2006}. As shown in
Fig.~\ref{cubicDepth}(b), the interference peaks of both scalar
and spinor BECs reversibly revive after the ramp-down sequences,
indicating atoms quickly recohere and return to SF states.

Observations in Fig.~\ref{cubicDepth} are qualitatively consistent with our MF calculations and suggest the existence of first-order SF-MI transitions under some circumstances. First,
LP spinor BECs at high magnetic fields possess many properties
(e.g., the peak OD) that are similar to those of scalar BECs.
Their ramp-up and ramp-down curves are close to each other, while
both have roughly symmetric transition points $u_{c}$. Similar
phenomena were observed in $^{87}$Rb and $^{6}$Li systems, and
have been considered as signatures of second-order SF-MI
transitions~\cite{Greiner02, LatticeRMP, Ketterle2006}. Second, LP
states at low magnetic fields and TP states at high fields
apparently have smaller $u_{c}$ for both ramp-up and ramp-down
processes compared to scalar BECs, suggesting enlarged Mott lobes.
Particularly, the ramp-down $u_{c}$ for LP states at low fields is
noticeably smaller than their ramp-up $u_c$, corroborating with the
MF picture that hysteresis occurs across the first-order phase
transitions. Third, the recovered interference contrast is visibly different for
various BECs after the ramp-down process (after SF-MI phase
transitions). For scalar and high-field LP spinor BECs, nearly
$75\%$ of peak OD can be recovered in the interference peaks after
the ramp-down sequence. The slightly reduced interference contrast
may be due to unaccounted heatings, which leads a small portion of
atoms ($< 20\%$) to populate the Brillouin zone. In contrast,
after we utilized quite a few techniques and optimized many
parameters, the maximal recovered interference contrast of
low-field LP states is only $\sim 40\%$ ($\sim 20\%$ for high-field
TP states). We attribute this to unavoidable heatings across the
first-order transitions as there is a jump in system energy between
meta-stable states and stable states. Both hysteresis effect and
significant heatings strongly suggest that first-order SF-MI
transitions are realized in our experiment. Note, however, we do
not see noticeable jumps in the observables as is typically
associated with first-order transitions. This is likely due to the
presence of even and odd atom fillings in inhomogeneous systems such
as trapped BECs, although the predicted first-order SF-MI transitions
only exist for even occupancy number. Limited experimental resolutions
may be another reason.

In addition, our data of the LP state in Fig.~\ref{cubicDepth}(b)
demonstrate the feasibility of realizing SF-MI transitions via a
new approach, i.e., by ramping $q_B$ at a fixed lattice depth. For example,
when the final $u_L$ in the ramp-down sequence is set at a value
between $17E_R$ and $21E_R$, atoms in the LP spinor BECs can cross
the SF-MI transitions if $q_B$ is sufficiently reduced (e.g., from
$h\times360\,$Hz to $h\times20\,$Hz). This agrees with the MF
prediction in Fig.~\ref{theory}(d): $u_c$ depends on $q_B$ in
antiferromagnetic spinor BECs.

We then compare scalar and spinor BECs within a wide range of
magnetic fields, $20 \, \text{Hz} \leq q_{B}/h \leq 500 \,
\text{Hz}$, after identical lattice ramp sequences to $u_L=10E_R$. We
choose $10E_R$ because it is apparently the lattice depth around which
we observe the maximum interference contrast, with negligible
difference in scalar and spinor BECs after the ramp-up sequence
at all $q_{B}$. This is consistent with Fig.~\ref{theory}, which
predicts all BECs studied in this work should be well in the SF phase
at $10E_R$. However, the interference peak ODs show intriguing
differences after the ramp-down sequence to $10E_R$ (see
Fig.~\ref{cubicQB}): deviations
from the maximal value appear for LP spinor BECs at low magnetic
fields and the TP state at all positive $q_{B}$. 
We again attribute this to different amount of
heatings across the SF-MI transitions. Different extent of heatings may be
produced due to different spin barriers as well as the amount of
energy jump across the transitions. Hence, the maximum recovered OD is
a good indicator for the appearance/disappearance of first-order phase
transitions. Notably, LP spinor BECs are found to behave very
similarly to scalar BECs as long as $q_B$ is large enough, i.e., $q_B
\geq h \times 100 \, \text{Hz} >U_{2}$ as shown in Fig.~\ref{cubicQB}.
This observation is again consistent with Fig.~\ref{theory}(d),  in which the
two MF curves for the LP state merge indicating that meta-stable
states disappear and SF-MI transitions become second order when
$q_{B}/h>70 \,$Hz. Furthermore, the
difference between LP and TP spinor BECs appears to exponentially
decrease as $q_B$ approaches zero. Exponential fits to the data verify that the LP and TP spinor BECs should show
the same behavior at $q_B=0$.

In conclusion, we have conducted the first experimental study on
the SF-MI phase transitions in an antiferromagnetic sodium spinor
BEC confined by 3D optical lattices. We have observed the
hysteresis effect and significant heatings across the phase
transitions, which suggest first-order SF-MI transitions are
realized in our experiment. These observations and the dependence
of the phase transitions on $q_B$ can be qualitatively understood
by MF theory. Further studies are required to confirm more
signatures of the first-order transitions, for example by
precisely imaging Mott shells~\cite{Chin2010, Campbell}. Our data
also suggest the feasibility of realizing SF-MI phase transitions
via changing the quadratic Zeeman energy.

\begin{acknowledgments}
We thank the ARO and the NSF for financial support. STW and LMD are supported by the
IARPA, the ARL, and the AFOSR MURI program. STW thanks Xiaopeng Li for
helpful discussions.
\end{acknowledgments}


\begin{thebibliography}{24}
\bibitem{spinOperator} The linear Zeeman energy is ignored in Eq.~\eqref{Eq:H}, since it remains the same during collisional spin interconversions in $F=1$ spinor BECs. The spin operators at site $i$ are given by %$S_{i\gamma=x,y,z}= b_{ik}^{\dagger}(F_{\gamma})_{kl}b_{il}$, where $F_{\gamma}$ is the $\gamma$ component of the spin-1 matrix.
$S_{ix} = \frac{1}{\sqrt{2}} \left( b_{i,0}^{\dagger}b_{i,1} +
b_{i,1}^{\dagger}b_{i,0}  + b_{i,-1}^{\dagger}b_{i,0} +
b_{i,0}^{\dagger}b_{i,-1} \right)$, $S_{iy} = \frac{i}{\sqrt{2}}
\left( b_{i,0}^{\dagger}b_{i,1} - b_{i,1}^{\dagger}b_{i,0}  +
b_{i,-1}^{\dagger}b_{i,0} - b_{i,0}^{\dagger}b_{i,-1} \right)$,
$S_{iz} = b_{i,1}^{\dagger}b_{i,1} - b_{i,-1}^{\dagger}b_{i,-1}$.
\bibitem{ramp} Lattices are linearly ramped up to a
given $u_L$ in a ramp-up sequence, while lattices are first
adiabatically ramped up to $26E_R$ and then back down to a
variable final $u_L$ in a ramp-down sequence. We find that a ramp
speed of $2E_R/\rm{ms}$ is sufficient to satisfy the intraband
adiabaticity condition and ensure $\geq 80\%$ of atoms remain in a
scalar or a high-field LP spinor BEC after a ramp-down sequence to $2E_R$.
\bibitem{KD} The 3D optical lattice is
constructed by three optical standing waves from a single-mode
laser at  $1064\,$nm, which results in a cubic periodic potential
with a lattice spacing of $532\,$nm. All lattice beams are
frequency-shifted by at least $20\,$MHz with respect to each other
for eliminating cross interference among them. The calibration of
$u_L$ is conducted via Kapitza-Dirac diffraction patterns and has
an uncertainty of $\sim15\%$.
\end{thebibliography}
\end{document}